\documentclass{pazh_engl}

\usepackage{graphicx}

\begin{document}

\title{\bf Hard X-ray sky survey with the SIGMA telescope aboard GRANAT observatory}

\author{\bf \hspace{-1.3cm}\copyright\, 2004 Ç. \ \
 M.G.Revnivtsev\inst{1,2}, R.A. Sunyaev\inst{1,2}, M.R.Gilfanov \inst{1,2},
 E.M.Churazov \inst{1,2}, A. Goldwurm\inst{3,4}, J.Paul\inst{3,4},
P.Mandrou\inst{5}, J.P.Roques\inst{5}}
\institute{Space Research Institute, Moscow, Russia
\and
Max-Planck-Institut fuer Astrophysik, Garching, Germany 
\and
DAPNIA/Service d'Astrophysique CEA-Saclay, France
\and
Fe'de'ration de Recherche Astroparticule
   et Cosmologie Universite' de Paris, France
\and
Centre d'Etude Spatiale des Rayonnements, Toulouse, France
} 
\authorrunning{Revnivtsev et al.}
\titlerunning{GRANAT/SIGMA survey of hard X-ray sky}
\date{Feb.19, 2004}
\abstract{
During the lifetime of GRANAT orbital observatory the SIGMA telescope
collected X-ray images of more than 1/4 of the whole sky. Among these
regions the Galactic Center had largest exposure time 
($\sim$9 million sec). In the present work we review all observations of the
SIGMA telescope and present sensitivities achieved with it at different
sky regions
}
\maketitle

\section*{INTRODUCTION}

SIGMA telescope aboard GRANAT observatory was the first space 
telescope that used coded aperture technique for reconstruction of 
sky images in hard X-rays (energy band 35-1300 keV). Large lifetime of the
telescope (it worked with some interruptions practically 8 years)
allowed us to obtain unique set of hard X-ray images of
the sky with unprecedented angular resolution ($\sim 15^\prime$) and
accuracy of a source localization ($\sim2-3^\prime$).

Because of these advantages, in particular, telescope SIGMA discovered
very interesting hard X-ray source GRS 1758-258, which is located 
only $40^\prime$ apart from bright soft X-ray source GX 5-1 (Mandrou
et al. 1991, Sunyaev et al. 1991). Large variations of hard X-ray flux 
($>$400 keV) was detected in the spectrum of black hole candidate 
1E1740.7-2942 (Bouchet et al. 1991, Sunyaev et al. 1991b).
There was detected hard X-ray flux
from X-ray burster A1742-294, which is very near to bright black hole
binary 1E1740.7-2942 (Churazov et al. 1995). There was put an upper
limit on hard X-ray flux of central supermassive black hole of our Galaxy 
(Sunyaev et al. 1991, Goldwurm et al. 1994)

In coming years one can expect a great step in the surveys of hard
X-ray sky (20-200 keV). In particular, new hard X-ray observatory
INTEGRAL (see Winkler et al.2003) was launched in 2002. It demonstrated
the ability to provide 5-8 time better sensitivity than it was available for
the SIGMA (see e.g. Revnivtsev et al. 2004). Another observatory with
hard X-ray coded mask telescope -- SWIFT (see e.g. Gehrels 2000) -- 
is planned for the launch in 2004. 
In addition to coded mask telescopes the reflection telescopes that
can work in the energy band 20-70 keV is being developed (e.g. SIMBOL-X, 
Fernando et al. 2002). In the view of these advances we decided to
publish the overview  of survey of hard X-ray sky of the 
the SIGMA telescope of GRANAT observatory.

During the period 1990-1998 the SIGMA observed more that one quarter of the sky
with sensitivity better than 100 mCrab. Galactic Center region
had the deepest exposure time ($\sim$9 million sec), providing the
sensitivity to a source discovery (5$\sigma$) approximately 10 mCrab.
In the present work we review observations of the telescope,
achieved sensitivities and the list of sources that were detected 
during the period 1990-1998.

\begin{figure*}[htb]
\includegraphics[height=\textwidth,angle=-90]{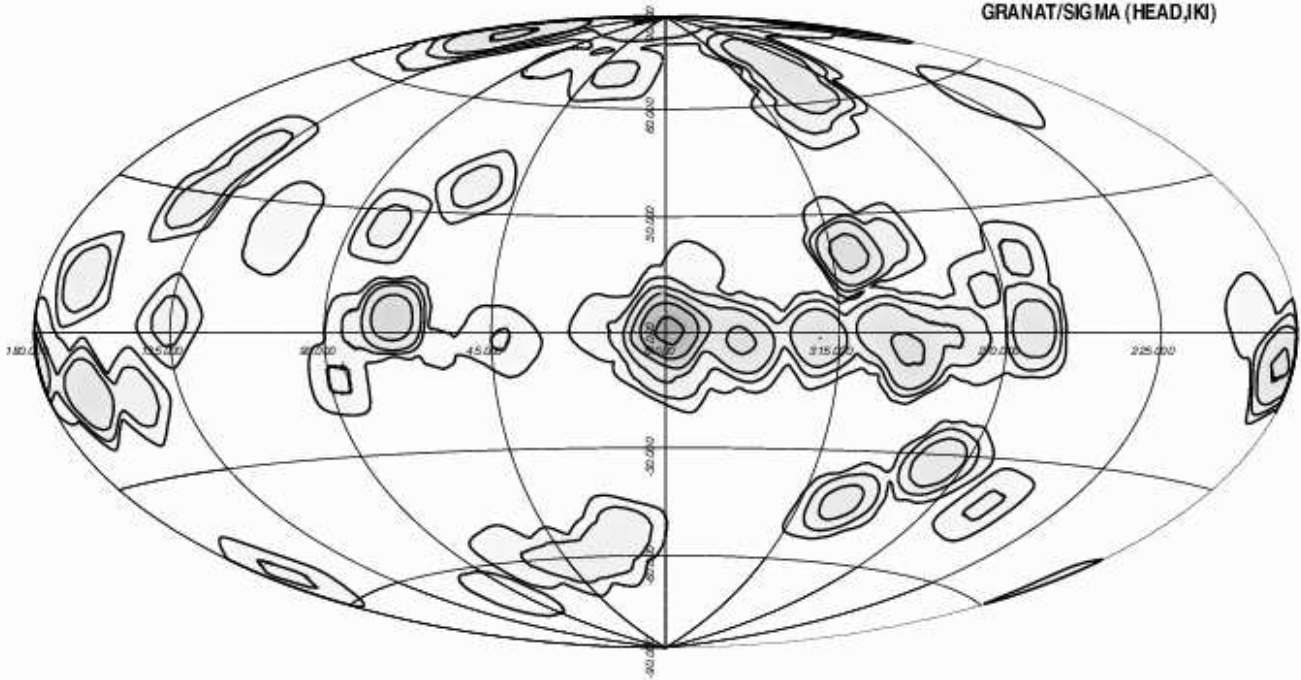}
\caption{Exposure map of observations of SIGMA telescope during period 1990-1998.
Countours denote region where exposure time is larger than 10,100, 315 ksec,
1.0, 3.16 and 8 Msec.}
\end{figure*}

\begin{figure}[htb]
\includegraphics[width=\columnwidth]{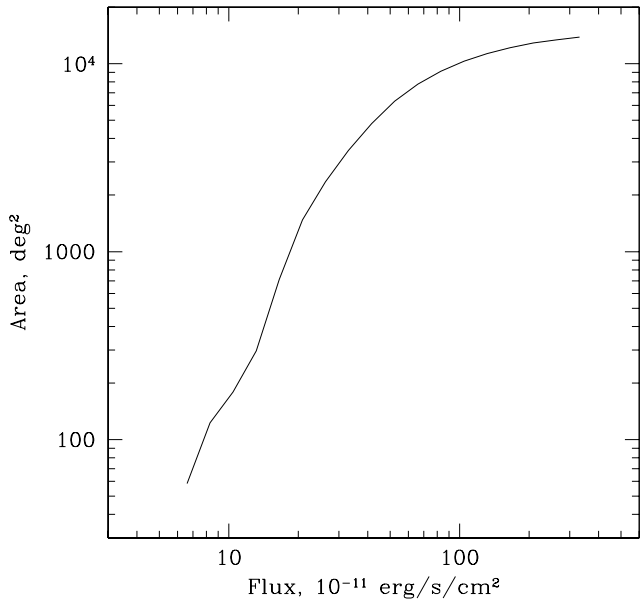}
\caption{Dependence of the solid angle of the sky covered with some sensitivity 
on the value of this sensityivity (5$\sigma$).
}
\end{figure}

\section*{RESULTS}

Joint French-Soviet hard X-ray telescope SIGMA was one of the main 
instruments aboard GRANAT orbital observatory. Coded aperture --
tungsten mask placed at 2.5m of the positional sensitive NaI detector --
allowed to reconstruct images of hard X-ray sky in the energy band 
35-1300 keV with angular resolution 
$\sim 15^\prime$.  Thanks to high apogee orbit of GRANAT (period of 
revolution $\sim$3 days, hight of apogee $\sim$200\,000 km) instruments 
of the observatory could have almost continuous observations during 3 days, 
with short interruptions for the telemetry dumps. Detailed description
of the SIGMA telescope can be found in Paul et al. (1991). In flight
performance of the telescope - in work of Mandrou et al. (1991)

During the period 1990-1998 the SIGMA performed more than 500 pointed
observations of different astrophysical objects with typical 
exposure times around 20-24 hours. Total exposure time
of all observations (corrected for deadtime fraction) is around
30 million sec. Of this 9 million of sec was spent to observe the 
Galactic Center region. Map of effective exposure (corrected for deadtime 
fraction 
and vignetting -- dependence of effective exposure on distance from 
the center of the field of view) of all performed observations
is presented in Fig.1. Contours on Fig.1 denote regions
which have exposure times more than 10, 100, 316 ksec, 1.0, 3.16 and 8 Msec.

40-100 keV energy band of the SIGMA telescope is the most sensitive to
typical sky X-ray objects. Dependence of sky solid angle on
sensitivity achieved by the SIGMA in the energy band 40-100 keV
is presented in Fig.2. Contours of sensitivity, achieved 
by the SIGMA in the Galactic Center region and region of Norma spiral arm 
tangent are presented in Fig.3. On Figures 2 and 3 sensitivities
correspond to 5$\sigma$ level, i.e. to the level of detection 
of unknown source. In order to obtain an upper limit on the hard X-ray flux
of a source with known position one can use slightly smaller limit, 
for example 2-3$\sigma$. In this case the best value would be $\sim 4-6$ 
mCrab. For a source with Crab-like spectrum 1mCrab flux value corresponds to
energy flux  $\sim10^{-11}$ ergs/s/cm$^2$ or
$\sim 10^{-5}$ phot/s/cm$^2$, that in turn corresponds to the luminosity
$\sim8\times10^{35}$ ergs/s at the distance of the Galactic Center (8.5 kpc).

During its operation time the SIGMA telescope detected a number of 
hard X-ray sources: galactic compact objects with neutron stars and black 
holes,  X-ray Novae, active galactic nuclei. In Table 1 we present
the list of detected sources. In the last column of the table we present
references to works where detailed information about the SIGMA results
on these sources can be found.

\begin{figure}[htb]
\includegraphics[width=\columnwidth]{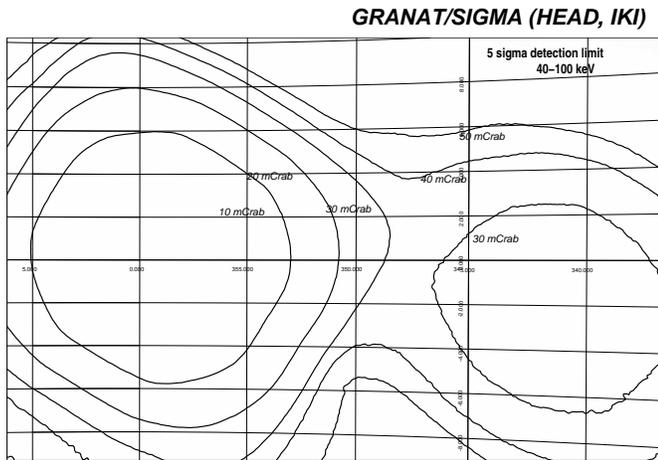}
\caption{Countours of sensitivity, achieved by the SIGMA for ne sources detection 
(5$\sigma$) in the region of Galactic Center and Norma arm tangent.
}
\end{figure}

Map of the whole sky, obtained by averaging of all observations is shown
in Fig.4. Zoomed image of the Galactic Center region and region of
Norma spiral arm tangent is presented in Fig.5. It is worth to note
that because we present here only averaged images of regions, some
weak and variable or transient sources are not visible on these images (
for example KS/GRS 1730-312, GRS 1739-278, KS 1731-260 and so on)

Typical sensitivity of the SIGMA for detection of new sources
over approximately one quarter of the sky is $\sim$100 mCrab, that is
somewhat worse that the previous all sky survey performed by 
scanning collimator A4 of HEAO1 observatory (typical sensitivity of
this survey in 40-100 keV energy band is $\sim$ 10--15 mCrab, see. Levine
et al. 1984). However, very good spatial resolution of the SIGMA
allowed it to make a set of interesting discoveries in a very crowded
region of the Galactic Center, that was impossible for HEAO1/A4.
Sensitivity of the SIGMA in the Galactic Center region is
$\sim$8--10 mCrab (5$\sigma$), that is comparable with that of HEAO1/A4.

\bigskip

Work is partially supported by grabt of Minpromnauka (grabnt of president of Russian Federation  NSH-2083.2003.2) and program of RAS ''Non stationary phenomena i astronomy''

\section*{REFERENCES}

L. Bassani, E. Jourdain, J.P. Roques et al., Astrophys.J. {\bf 396}, 504 (1992)

L. Bassani, E. Jourdain, J.P. Roques et al., Astron.Astrophys.Suppl.Ser. {\bf 97}, 89 (1993)

D. Barret, S. Mereghetti, J.P. Roques et al., Astrophys.J. {\bf 379}, 21 (1991)

D. Barret, J.P. Roques, P. Mandrou et al., Astrophys.J. {\bf 392}, 19 (1992)

D. Barret, P. Mandrou, J.P. Roques et al., Astron.Astrophys.Suppl.Ser. {\bf 97}, 241 (1993)

G. Belanger, A. Goldwurm, P. Goldoni et al., Astroph.J. {\bf 601}, 163L (2004)

I.A. Bond, J. Ballet, M. Denis  et al., Astron.Astrophys. {\bf 307}, 708 (1996)

V. Borrel, L. Bouchet, E. Jourdain et al., Astrophys.J. {\bf 462}, 754 (1996)

L. Bouchet, P. Mandrou, J.P. Roques et al., Astrophys.J. {\bf 383}, 45L (1991)

L. Bouchet, E. Jourdain, P. Mandrou et al., Astrophys.J {\bf 407}, 739 (1993)

E. Churazov, M. Gilfanov, R. Sunyaev et al., IAU Circ. 5623

E. Churazov, M. Gilfanov, R. Sunyaev et al., Astrophys.J.Suppl.{\bf 92}, 381 (1994a)  

E. Churazov, M. Gilfanov, A. Finoguenov et al., IAU Symposium {\bf 159}, 63-72 (1994b)

E. Churazov, M. Gilfanov, R. Sunyaev et al., Astrophys.J. {\bf 443}, 341 (1995)

A. Claret, J. Ballet, A. Goldwurm et al., Adv.Space Res. {\bf 13}, 735 (1993)

A. Claret, A. Goldwurm, B. Cordier et al., Astrophys.J. {\bf 423}, 436 (1994)

B. Cordier, A. Goldwurm, P. Laurent et al., Adv.Space Res. {\bf 11}, 169 (1991)

P. David, P. Laurent, M. Denis et al., Astron.Astrophys. {\bf 332}, 165 (1998)

M. Denis, J.P. Roques, D. Barret et al., Astron.Astrophys.Suppl.Ser. {\bf 97}, 333 (1993)

M. Denis, J. Olive, P. Mandrou et al., Astrophys.J.Suppl. {\bf 92}, 459 (1994)

Ph. Ferrando, Proceedings of the 4th Microquasar Workshop, eds. Ph Durouchoux, Y. Fuchs and J. Rodriguez, Corsica, France, 2002, astro-ph/0209062 (2002)

A. Finoguenov, E. Churazov, M. Gilfanov et al., Astrophys.J {\bf 424}, 940 (1994)

A. Finoguenov, E. Churazov, M. Gilfanov et al., Astron.Astrophys. {\bf 300}, 101 (1995)

A. Finoguenov, M. Gilfanov, E. Churazov et al., Astron.Lett. {22}, 721 (1996)

N. Gehrels N., Proc. SPIE Vol. 4140, 42 (2000)

M. Gilfanov, R. Sunyaev, E. Churazov et al., Sov.Astron.Lett. {\bf 17}, 437 (1991)

M. Gilfanov, E. Churazov, R. Sunyaev et al., Astron.Astrophys. {\bf 418}, 844 (1993)

M. Gilfanov, E. Churazov, R. Sunyaev et al., Astrophys.J.Suppl.Ser., {\bf 92}, 411 (1994)

A. Goldwurm, J. Ballet, B. Cordier et al., Astrophys.J. {\bf 389}, 79 (1992)

A. Goldwurm, B. Cordier, J. Paul et al., Nature {\bf 371}, 589 (1994)

A. Goldwurm, M. Vargas, J. Paul et al., Astron.Astrophys. {\bf 310}, 857 (1996)

P. Goldoni, M. Vargas, A. Goldwurm et al., Astron.Astrophys. {\bf 239}, 186 (1998)

P. Goldoni, M. Vargas, A. Goldwurm et al., Astrophys.J. {\bf 511}, 847 (1999)

E. Jourdain, L. Bassani, J.P. Roques et al., Astrophys.J. {\bf 395}, 69 (1992a)

E. Jourdain, L. Bassani, L. Bouchet et al., Astron.Astrophys. {\bf 256}, 38 (1992b)

S. Kuznetsov, M. Gilfanov, E. Churazov et al., MNRAS {\bf 292}, 651 (1997)

S. Kuznetsov, M. Gilfanov, E. Churazov et al., Astron.Lett {\bf 25}, 351 (1999)

F. Lebrun, J. Ballet, J. Paul et al., Astron.Astrophys. {\bf 264}, 22 (1992)

A. Levine, F. Lang, W. Lewin et al., Astrophys.J.Suppl.Ser. {\bf 54}, 581 (1984)

P. Laurent, A. Goldwurm, F. Lebrun et al., Astron.Astrophys. {\bf 260}, 237 (1992)

P. Laurent, B. Cordier, A. Goldwurm et al., Adv.Space Res. {\bf 13}, 751 (1993a)

P. Laurent, L. Salotti, J. Paul et al., Astron.Astrophys. {\bf 278}, 444 (1993b)

P. Laurent, J. Paul, A. Claret et al., Astron.Astrophys.,{\bf 286}, 838 (1994)

P. Laurent, J. Paul, M. Denis et al.,Astron.Astrophys. {\bf 300}, 399 (1995)

P. Mandrou, J.P. Chabaud and M. Ehanno, Gamma-ray line astrophysics, Proceedings of the International Symposium, Paris, France, Dec. 10-13, 1990, New York, American Institute of Physics, 1991, 492

S. Mereghetti, J. Ballet, A. Lambert et al., Astrophys.J {\bf 366}, 23 (1991)

J. Meji'a, T. Villela, P. Goldoni et al., Astrophys.J. {\bf 566}, 387 (2002)

M. Pavlinsky, S. Grebenev, R. Sunyaev, Sv.Astr.Lett  {\bf 18}, 291 (1992)

M. Pavlinsky, S. Grebenev, R. Sunyaev, Astrophys.J. {\bf 425
}, 110 (1994)

J. Paul, J. Ballet, M. Cantin et al., Adv.Space Res. 11, 279

í. Revnivtsev, í. Gilfanov, å. Churazov et al., Joint European and National Astronomical Meeting JENAM-97,  ôhessaloniki, Greece, p.289 (1997)

M. Revnivtsev, M. Gilfanov, E. Churazov et al., Astron.Astrophys. {\bf 331}, 557 (1998)

M. Revnivtsev, M. Gilfanov, E. Churazov et al., Astron.Lett. {\bf 25}, 493 (1999)

J.P. Roques, L. Bouchet, E. Jourdain et al., Astrophys.J.Suppl. {\bf 92}, 451 (1994)

L. Salotti, J. Ballet, B. Cordier et al., Astron.Astrophys. {\bf 253}, 145 (1992)

A. Sitdikov, M. Gilfanov, R. Syunyaev et al., Astron.Lett. {\bf 19}, 188 (1993)

R. Sunyaev, E. Churazov, V. Efremov, M. Gilfanov, S. Grebenev, Adv.Space Res. {\bf 10}, 41 (1990)

R. Sunyaev, E. Churazov, M. Gilfanov et al., Astron. Astrophys. {\bf 247}, 29 (1991a)

R. Sunyaev, E. Churazov, M. Gilfanov et al., Astrophys.J. {\bf 383}, 49L (1991b)

R. Sunyaev, E. Churazov, M. Gilfanov et al., Astrophys.J. {\bf 389}, 75 (1992)

R. Sunyaev, S. Grebenev, A. Lutovinov et al., Astronomer's Telegram 190 (2003)

S. Trudolyubov, M. Gilfanov, E. Churazov et al., Astron.Lett. {\bf 22}, 664 (1996)

S. Trudolyubov, M. Gilfanov, E. Churazov et al., Astron.Astrophys. {\bf 334}, 895 (1998)

S. Trudolyubov, E. Churazov, M. Gilfanov et al., Astron.Astrophys. {\bf 342}, 496 (1999)

M. Vargas, A. Goldwurm, M. Denis et al., Astron.Astrophys.Suppl. {\bf 120}, 291 (1996)

M. Vargas, A. Goldwurm, P. Laurent et al., Astrophys.J. {\bf 476}, 23 (1997) 

A. Vikhlinin, E. Churazov, M. Gilfanov et al., Astrophys.J. {\bf 424}, 395 (1994)

A. Vikhlinin, E. Churazov, M. Gilfanov et al., Astrophys.J. {\bf 441}, 779 (1995)

C. Winkler, T.J.-L. Courvoisier, G. Di Cocco et al., Astron. Astrophys. {\bf 411}, 1 (2003)

\begin{table*}
{{\bf Table 1.}. Table of sources detected by the SIGMA in energy band 40-100 keV}

\footnotesize
\begin{tabular}{l|c|l|l}
Source& $F_{\rm max,40-100 keV}$, mCrab&Class& Refs\\
\hline
\multicolumn{3}{c}{Galactic sources}\\
\hline
Crab$^h$&1000&Pulsar&1\\
Cyg X-1$^h$&2000&BH&2,3,4\\
1E1740.4-2942$^h$&150&BH&4,5,6,7\\
GRS 1758-258$^h$&100&BH&5,8,9,10\\
GRS 1915+105$^h$&150&BH&11\\
GX339-4$^h$&430&BH&12,13\\
SLX 1735-269&20&NS, Burster&14\\
KS 1731-260&70&NS, Burster&15\\
TrA X-1&80&BH?&16\\
4U1724-30(Terzan 2)&50&NS, Burster&16,17\\
H1732-304(Terzan 1)&10&NS, Burster&18\\
GX354-0(4U1728-34)&100&NS, Burster&19\\
A1742-294&30&NS, Burster&20\\
4U1705-44&70&NS, Burster&21\\
4U1608-52&70&NS, Burster&21\\
4U1700-37&200&NS,HMXB&22,23\\
OAO 1657-415&100&NS, Accr.Pulsar&24\\
Vela X-1&200&NS, Accr.Pulsar&25\\
GX1+4&100&NS, Accr.Pulsar&26,27\\
PSR 1509-58&17&NS, Pulsar&28\\
GRS 0834-430&100&NS, Accr.Pulsar&29\\
GRO J1744-28&120&NS, Accr.Burster-pulsar&30\\
Sgr A$^*$&$<$10&SBH&5,31\\
GRS 1227+025&100&&32\\
\hline
\multicolumn{3}{c}{Extragalactic sources}\\
\hline
Cen A&130&AGN&33,34,35\\
3C273&40&Blasar&34,36\\
NGC 4151&40&AGN&34,37,38\\
NGC 4388&10&AGN&39\\
GRS 1734-292&36&AGN&40\\
\hline
\multicolumn{3}{c}{X-ray Novae}\\
\hline
Nova Musca 91 (GS/GRS 1124-68)&1070&BH&41,42,43\\
Nova Persei 92 (GRO J0422+32)&2800&BH&44,45,46,47,48\\
Nova Oph 93 (GRS 1716-249)&1200&BH&49\\
Nova Vel 93 (GRS 1009-45)&65&BH&50\\
KS/GRS 1730-312&170&BH&51,52\\
GRS 1739-278&100&BH&53\\
GRS 1737-31&115&BH&54\\
XTE J1755-324&85&BH&55,56\\
\end{tabular}
{\footnotesize
\begin{list}{}
\item $^h$ -  source also was detected in 100--200 keV energy band

\item  1- Gilfanov et al. 1994,
2 - Salotti  et al. 1992,
3 - Vikhlinin et al. 1994,
4 - Kuznetsov et al. 1997,
5 - Sunyaev et al. 1991a,
6 - Bouchet et al. 1991,
7 - Sunyaev et al. 1991b,
8 - Laurent et al. 1993a,
9 - Gilfanov et al. 1993,
10 - Kuznetsov et al. 1999,
11 - Finoguenov et al. 1994,
12 - Bouchet et al. 1993,
13 - Trudolyubov et al. 1998, 
14 - Goldwurm et al. 1996,
15 - Barret et al. 1993,
16 - Barret et al. 1992, 
17 - Barret et al. 1991, 
18 - Borrel et al. 1996,
19 - Claret et al. 1994, 
20 - Churazov et al. 1995, 
21 - Revnivtsev et al. 1997, 
22 - Laurent et al. 1992, 
23 - Sitdikov et al. 1993, 
24 - Mereghetti et al. 1991, 
25 - Laurent et al. 1995, 
26 - Laurent et al. 1993b,
27 - David et al. 1998, 
28 - Laurent et al. 1994,
29 - Denis et al. 1993,
30 - Meji'a et al. 2002,
31 - Goldwurm et al. 1994,
32 - Jourdain et al. 1992a,
33 - Bassani et al. 1993, 
34 - Churazov et al. 1994b,
35 - Bond et al. 1996,
36 - Bassani et al. 1992, 
37 - Jourdain et al. 1992b,
38 - Finoguenov et al. 1995,
39 - Lebrun et al. 1992,
40 - Churazov et al. 1992,
40 - Gilfanov et al. 1991,
41 - Sunyaev et al. 1992, 
42 - Goldwurm et al. 1992,
43 - Claret et al. 1993,
44 - Roques et al. 1994,
45 - Denis et al. 1994,
46 - Vikhlinin et al. 1995,
47 - Finoguenov et al. 1996,
48 - Revnivtsev et al. 1998,
49 - Goldoni et al. 1998,
50 - Trudolyubov et al. 1996,
51 - Vargas et al. 1996, 
52 - Vargas et al. 1997,
53 - Trudolyubov et al. 1999,
54 - Revnivtsev et al. 1999,
55 - Goldoni et al. 1999
\end{list}
}
\end{table*}

\clearpage

\begin{figure*}
\includegraphics[width=\textwidth]{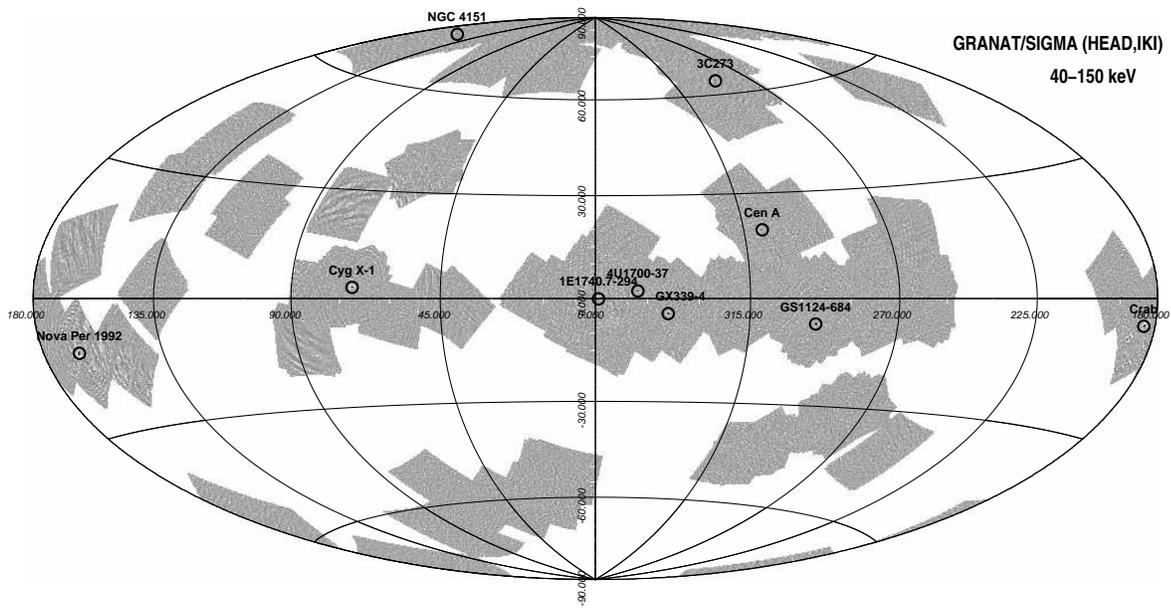}
\caption{Map of the whole sky averaged over all observations. Only brightest sources
are marked.}
\end{figure*}

\begin{figure*}
\includegraphics[height=\textwidth,angle=-90]{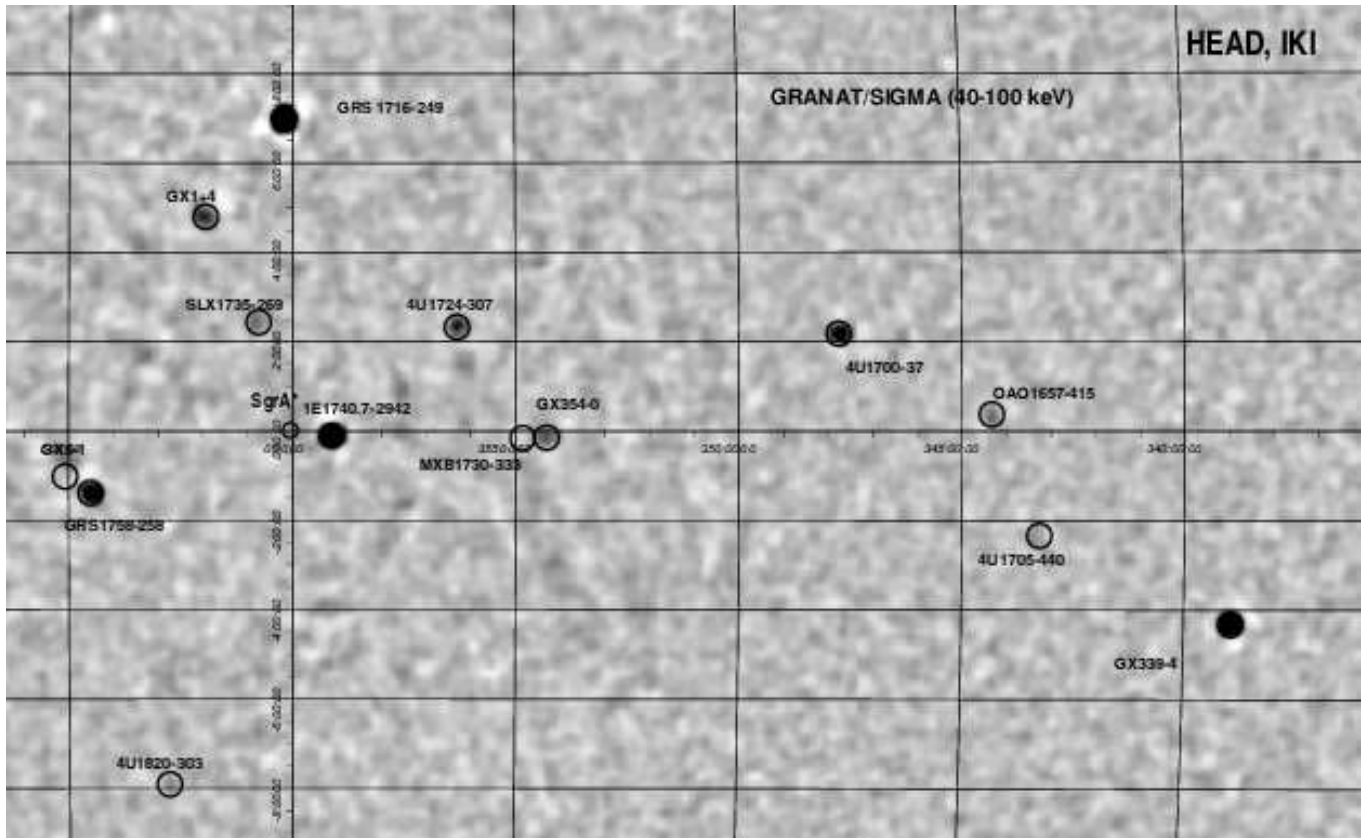}
\caption{Map of the Galactic Center region, obtained by averaging of all SIGMA 
observations.
}
\end{figure*}

\end{document}